\RequirePackage{fix-cm}
\documentclass[smallextended]{svjour3}       
\smartqed  
\usepackage{cite}
\usepackage{graphicx}
\usepackage{braket}
\usepackage{amsmath,amssymb}
\usepackage{amsfonts}
\usepackage{graphicx}
\usepackage{comment}
\usepackage{graphicx}
\usepackage{epsfig}
\usepackage{dcolumn}
\usepackage{subcaption} 
\usepackage{adjustbox}
\usepackage{float}
\usepackage{dsfont}	
\usepackage{relsize}
\usepackage{verbatim}
\usepackage{float}
\usepackage{ragged2e}
\usepackage{braket}
\usepackage{color}
\usepackage{xcolor}
\usepackage[ruled,vlined]{algorithm2e}

\SetCommentSty{mycommfont}

\usepackage{tikz}%
\usepackage{quantikz}%
\usepackage{float}
\usepackage{rotating}
\usepackage{subcaption}
\usepackage{graphicx}      
\usepackage[table,xcdraw]{xcolor}  
\usepackage{multirow}      

\usepackage{booktabs, multirow, caption}
\usepackage{amsmath}

\usepackage[hyphens]{url}  
\usepackage{graphicx} 
\usepackage{caption} 
\usepackage{algorithmic}

\begin{document}

\title{Low overhead circuit cutting with operator backpropagation
}


\author{Debarthi Pal \and Ritajit Majumdar
}


\institute{Debarthi Pal \at
            Indian Institute of Science, Bangalore\\
            \email{debarthipal@iisc.ac.in, paldebarthi@gmail.com}\\
            \emph{Debarthi Pal worked on this project during her summer internship at IBM Quantum}
            \and
            Ritajit Majumdar \at
            \emph{IBM Quantum}, IBM India Research Lab\\
            \email{majumdar.ritajit@ibm.com}
}


\maketitle

\begin{abstract}
    Current quantum computers suffer from noise due to lack of error correction. Several techniques to mitigate the effect of noise have been studied, in particular to extract the expectation value of observables. One such technique, circuit cutting, partitions large circuits into smaller, less noisy subcircuits, but the exponential increase in the number of circuit executions limits its scalability. Another method, operator backpropagation (OBP) reduces circuit depth by classically simulating parts of it, yet often escalates the number of circuit executions by some factor due to additional non-commuting terms in the updated observable. This paper introduces an optimized approach for minimizing noise in quantum circuits using operator backpropagation (OBP) combined with circuit cutting. We demonstrate that the strategic use of OBP with circuit cutting can mitigate the execution overhead. By employing simulated annealing, our proposed method identifies the optimal backpropagation parameter for specific circuits and observables, maximizing resource reduction in cutting. Results show a $3\times$ and $10\times$ decrease in resource requirements for Variational Quantum Eigensolver and Hamiltonian simulation circuits respectively, while maintaining or even enhancing accuracy. This approach also yields similar savings for other circuits from the Benchpress database and various observable weights, providing an efficient method to lower circuit cutting overhead without compromising performance.
\end{abstract}


\section{Introduction}
Quantum computers, despite their immense potential, are currently hindered by intrinsic system noise. To mitigate this challenge, three primary approaches have been identified: error correction in the long run and error suppression and mitigation in the near term. These near-term methods are mainly intended for problems that require the estimation of the expectation value of some observables \cite{di2024quantum, alexeev2024quantum, abbas2024challenges, basu2023towards}. Error mitigation techniques usually come at the expense of an increased number of circuit executions. 

One such method is called \emph{circuit cutting} \cite{peng2020simulating}, where a circuit can be partitioned into multiple smaller subcircuits by severing a wire (called wire cut henceforth) or replacing a 2-qubit gate with multiple instances of single-qubit gates or measurements (called gate cut henceforth). The general drawback of circuit cutting is that the number of circuit executions grow exponentially with the number of cuts. In particular, it has an overhead of $9^k$ for $k$ gate cuts, and $16^k$ for $k$ wire cuts. This makes its applicability limited to shallow circuits which can be partitioned into disjoint subcircuits using a small number of cuts only. A different technique, called \emph{Operator backpropagation} (OBP) \cite{fuller2025improved} reduces the depth of the circuit, at the expense of more circuit executions due to the increased number of non-commuting groups in the updated observable. The increase in the number of circuit executions caused by OBP is, in general, multiplicative. However, OBP can reduce the depth of the circuit, which often leads to lowering the number of cuts required to partition the circuit -- providing an exponential reduction in the number of circuit executions.

In this paper, we show that circuit cutting, equipped with OBP, can lead to $3-10\times$ reduction in the number of circuit executions when the backpropagation parameter is carefully selected. This significantly reduces the cutting overhead, making it applicable to larger and deeper circuits. We show such reductions for 6-qubit VQE, 19-qubit Hamiltonian simulation, and multiple other circuits from the Benchpress database \cite{nation2025benchmarking}. We execute the first two circuits and show that cutting with OBP can even provide improved quality of result, since the subcircuits now have even lower depth. We also show that it is not trivial to find the optimal parameter for OBP, and naive use of it can end up increasing the number of circuit executions instead of lowering it. For example, consider a circuit which has many layers of non-clifford gates towards the end. Here, the number of non-commuting groups in the observable will rapidly increase while backpropagating the non-clifford gates (see Sec.~\ref{sec:background} for more details). Naively choosing a small value for the backpropagation parameter may not be able to reduce the number of cuts, but will increase the number of circuit executions. In order to avoid such scenarios, we propose a Simulated Annealing-based method to find the optimal backpropagation parameter, while reducing the classical overhead of backpropagation by avoiding a brute force search over the parameter space.

The rest of the paper is organized as follows: In Sec.~\ref{sec:background} we give a brief overview of circuit cutting and operator backpropagation. Sec.~\ref{sec:obp_cut} formulates the problem of using OBP with cutting, and shows the reduction in the circuit executions for 6-qubit VQE, 19-qubit Hamiltonian simulation, as well as other circuits from Benchpress, and the improvement in the quality of outcome for the first two. Sec.~\ref{sec:select_qwc} discusses the issues of naively using OBP with cutting, and proposes the simulated annealing method for selection of the optimal backpropagation parameters. In Sec.~\ref{sec:truncation} we briefly touch upon OBP \emph{with truncation} for circuit cutting, and discuss similar potential benefit of using Multi-product formula (individually or together with OBP) for trotter circuits in Sec.~\ref{sec:discussion}. We conclude in Sec.~\ref{sec:conclusion}.

\section{Background}
\label{sec:background}
In this section we shall briefly discuss circuit cutting and operator backpropagation (OBP). The following section shall show how to reduce the overhead of circuit cutting using OBP.

\subsection{Circuit cutting}
\label{subsec:cut}
Circuit cutting is a method of partitioning a circuit into multiple smaller subcircuits either by \emph{cutting} along the wire, or replacing a two qubit gate by multiple single qubit gates and/or measurement. This method allows to (i) execute larger circuits on smaller hardware, and/or (ii) reduce the noise in the circuit execution since each subcircuit has fewer qubits and/or gates. Circuit cutting is mainly of two types:

\begin{itemize}
    \item Wire cutting \cite{peng2020simulating, tang2021cutqc, brenner2023optimal} is a decomposition method that partitions a quantum circuit into smaller, independent subcircuits by virtually severing wires between gates. The process involves measuring the output state of one subcircuit in a tomographically complete basis, and preparing a corresponding input for the following subcircuit in the corresponding eigenstate. In other words, given a quantum state $\rho$ and an observable $A$, $Tr\{\rho A\} = \sum_i c_i Tr\{AO_i\}\rho_i$, where $\{O_i\}$ is a tomographically complete basis (often taken as the Pauli basis), $\rho_i$ denote the corresponding eigenstate, and $c_i \in \{+1, -1\}$.

    \item Gate cutting \cite{mitarai2021constructing, mitarai2021overhead, schmitt2025cutting} decomposes a two-qubit gate into multiple single-qubit operations or projective measurements, each associated with a specific probability. In other words, a 2-qubit gate $U_{AB} = \sum_i a_i F_i^A \otimes F_i^B$, where $F_i$ indicate single qubit operation, or measurement, and $a_i$ forms a quasi-probability distribution, i.e. $\sum_i a_i = 1$. The exact values of $a_i$, and the corresponding operations $F_i$, depend on the type of two-qubit gate being cut. 
\end{itemize}


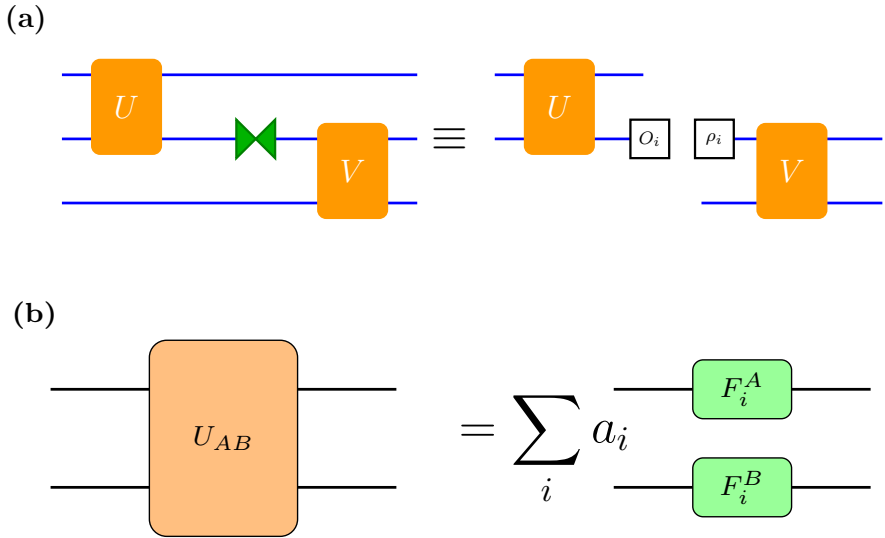
\begin{figure}[htb]

\begin{subfigure}[b]{0.75\columnwidth}
\centering


\scalebox{0.85}{
\begin{tikzpicture}[line width=1pt]
\node[anchor=north west] at (-1,3.2) {\LARGE\textbf{(a)}};

\draw[blue, very thick] (0,2) -- (5.5,2);
\draw[blue, very thick] (0,1) -- (5.5,1);
\draw[blue, very thick] (0,0) -- (5.5,0);

\node[fill=orange!80!yellow, rounded corners=4pt, minimum width=1.1cm, minimum height=1.5cm, text=white]
    at (1,1.5) {\Large $U$};
\node[fill=orange!80!yellow, rounded corners=4pt, minimum width=1.1cm, minimum height=1.5cm, text=white]
    at (4.5,0.5) {\Large $V$};

\filldraw[fill=green!70!black, draw=green!50!black]
    (2.7,1.3) -- (3.0,1.0) -- (2.7,0.7) -- cycle;
\filldraw[fill=green!70!black, draw=green!50!black]
    (3.3,1.3) -- (3.0,1.0) -- (3.3,0.7) -- cycle;

\node at (6.0,1) {\huge $\equiv$};

\draw[blue, very thick] (6.7,2) -- (9.0,2);
\draw[blue, very thick] (6.7,1) -- (8.8,1);
\node[fill=orange!80!yellow, rounded corners=4pt, minimum width=1.1cm, minimum height=1.5cm, text=white]
    at (7.7,1.5) {\Large $U$};

\node[draw=black, fill=white, minimum width=0.6cm, minimum height=0.6cm]
    at (9.1,1.0) {\small $O_i$};

\draw[blue, very thick] (9.9,1) -- (12.7,1);
\draw[blue, very thick] (9.9,0) -- (12.7,0);
\node[draw=black, fill=white, minimum width=0.6cm, minimum height=0.6cm]
    at (10.1,1.0) {\small $\rho_i$};
\node[fill=orange!80!yellow, rounded corners=4pt, minimum width=1.1cm, minimum height=1.5cm, text=white]
    at (11.3,0.5) {\Large $V$};
\end{tikzpicture}
}
\vspace{1em}
\end{subfigure}

\vspace{1.0em}

\begin{subfigure}[b]{0.6\columnwidth}
\centering


\scalebox{1.3}{
\begin{tikzpicture}
\node[anchor=north west] at (-1.5,2.5) {\textbf{(b)}};

\draw[fill=orange!50,rounded corners=5pt] (0,0) rectangle (1.5,2)
    node[pos=.5] {$U_{AB}$};
\draw[-,thick] (-1,1.5) -- (0,1.5);
\draw[-,thick] (-1,0.5) -- (0,0.5);
\draw[-,thick] (1.5,1.5) -- (2.5,1.5);
\draw[-,thick] (1.5,0.5) -- (2.5,0.5);

\node at (4.0,0.9) {\Large$\displaystyle = \sum_i a_i$};

\draw[fill=green!40,rounded corners=3pt] (5.5,1.2) rectangle (6.5,1.8)
    node[pos=.5] {$F_i^A$};
\draw[fill=green!40,rounded corners=3pt] (5.5,0.2) rectangle (6.5,0.8)
    node[pos=.5] {$F_i^B$};
\draw[-,thick] (4.7,1.5) -- (5.5,1.5);
\draw[-,thick] (4.7,0.5) -- (5.5,0.5);
\draw[-,thick] (6.5,1.5) -- (7.3,1.5);
\draw[-,thick] (6.5,0.5) -- (7.3,0.5);
\end{tikzpicture}
}
\vspace{1em}
\end{subfigure}

\caption{\textbf{(a)} A 3-qubit circuit with gates \( U \) and \( V \), and interaction on the middle qubit (green triangles), rewritten as two 2-qubit circuits with a virtual break between them. \textbf{(b)} Cutting a 2-qubit gate \(U_{AB}\) into multiple instances of single-qubit gates \(F_i^{A}\) and \(F_i^{B}\). The coefficients \(a_i\) form a quasi-probability distribution, i.e., \(\sum_i a_i = 1\), but it is not necessary that \(0 \leq a_i \leq 1\) $\forall$ \(i\).}
\label{fig:combined}
\end{figure}

In Fig.~\ref{fig:combined} (a) and (b) we show a representation of wire cutting and gate cutting respectively. In both of these methods, the constructed subcircuits can be executed independently of each other \cite{bhoumik2023distributed, khare2023parallelizing}, and the original expectation value of the required observable can be obtained by reconstructing the outcomes of the individual subcircuits classically. Since the reconstruction is classical, it is noise-free, and each subcircuit is expected to incur lesser noise. Several research have shown the usefulness of circuit cutting to reduce the effect of noise \cite{bhoumik2023distributed, khare2023parallelizing, majumdar2022error, basu2022qer, saleem2021divide}. However, the drawback of this method is that the number of circuit execution increases exponentially with the number of cuts ($16^k$ for $k$ wire cuts, and $9^k$ for $k$ gate cuts), making this method impractical for larger and deeper circuits.

\subsection{Operator backpropagation (OBP)}
\label{subsec:obp}
Operator backpropagation (OBP) \cite{fuller2025improved} is an error mitigation technique designed to reduce the impact of noise on quantum circuits by effectively lowering their depth. In OBP, instead of executing a deep quantum circuit directly, an observable is propagated backward through the circuit, resulting in a shallower equivalent circuit. 

\begin{figure}[htb]
\begin{center}
\includegraphics[scale=0.35]{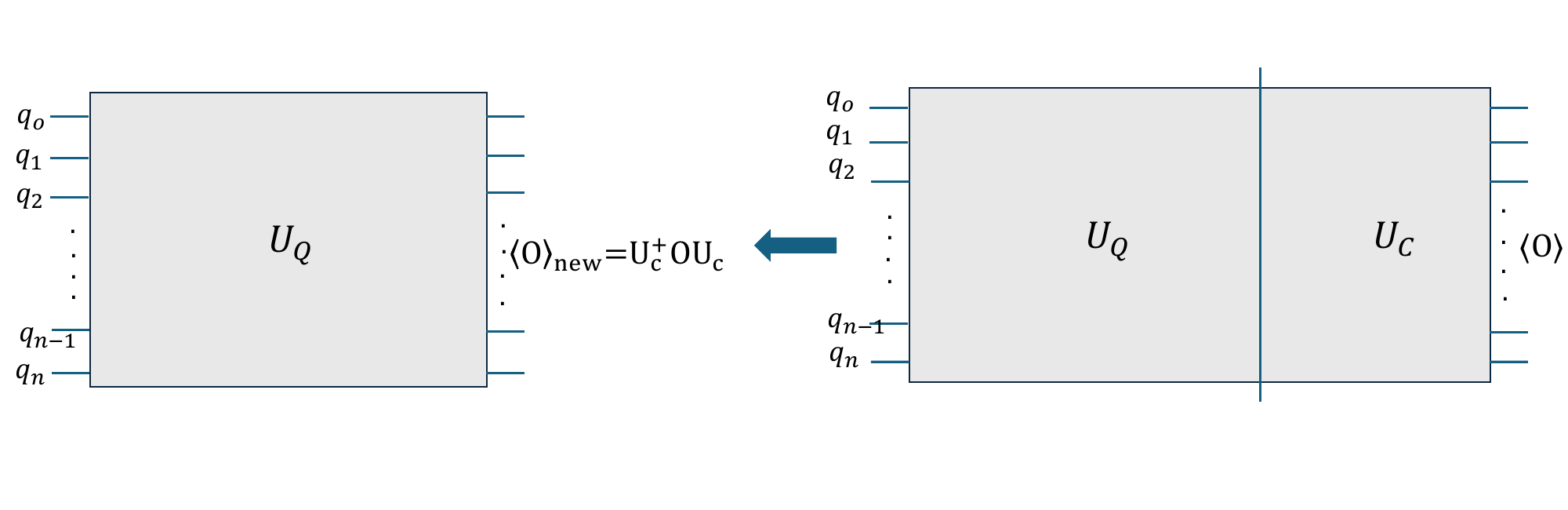}
\caption{A circuit unitary $U$, virtually divided into $U = U_CU_Q$ with $O$ as the observable whose expectation value is to be calculated. After backpropagation $U_CU_Q$ becomes $U_Q$ having lower depth and updated observable as $O_{new} = U_C^{\dagger}OU_C$}
\label{fig:obp_diag}
\end{center}
\end{figure}

Let $O$ be the original observable whose expectation value is to be calculated. If, $|\psi \rangle _{init}$ is the initial quantum state, and $|\psi \rangle$ be the final quantum state after execution of the circuit, then the expectation value is 
\begin{eqnarray}
    \langle \psi|O|\psi \rangle &=& _{init}\langle \psi|U^{\dagger}OU|\psi \rangle _{init} \nonumber \\
    &=& _{init}\langle \psi|U_Q^{\dagger}U_C^{\dagger}OU_CU_Q|\psi \rangle _{init} \nonumber \\
    &=& _{init}\langle \psi|U_Q^{\dagger}O_{new}U_Q|\psi \rangle _{init} \nonumber
\end{eqnarray}

where the original circuit unitary $U$ is \emph{logically} divided into $U = U_CU_Q$. The new circuit $U_Q$ has a lower depth, and the updated observable is $O_{new} = U_C^{\dagger}OU_C$. This method is shown pictorially in Fig.~\ref{fig:obp_diag}.

If the circuit is clifford, then it can be completely backpropagated without any increase in the number of non-commuting groups in the observable since a clifford always maps a Pauli to another Pauli. For a non-clifford gate, however, this may lead to an increased number of non-commuting terms in the new observable. Non-clifford gates can be represented as $U_j=e^{-\frac{i\theta_jP_j}{2}}$ where $P_j$ is a Pauli and $|\theta_j|\leq \frac{\pi}{4}$ \cite{beguvsic2025simulating}. Applying OBP changes the observable $O$ to

\begin{equation*}
    e^{i\theta_j P_j / 2} O e^{-i\theta_j P_j / 2} =
\begin{cases}
O, & [P_j, O] = 0, \\
\cos(\theta_j) O + i \sin(\theta_j) P_j O, & \{P_j, O\} = 0
\end{cases}
\end{equation*}

When two observables do not commute, separate quantum circuits are required to evaluate the expectation value for each, leading to an increased number of circuit executions. This introduces a trade-off: while OBP reduces the depth of the circuit, it can significantly increase the number of circuit executions, depending on the number of non-commuting observables produced. To manage this trade-off and control computational cost, OBP includes a parameter called max\_qwc\_groups. This parameter sets an upper limit on the number of groups of non-commuting observables—also known as qubit-wise commuting (QWC) groups—that will be allowed during the backpropagation process. Backpropagation will stop once the number of non-commuting group exceeds max\_qwc\_groups. By tuning max\_qwc\_groups, users can balance between reducing noise (via lower circuit depth) and minimizing the number of circuit executions required, thereby optimizing the overall resource usage within their computational budget.

\section{Reducing circuit cuts with OBP}
\label{sec:obp_cut}
 
In Sec.~\ref{sec:background} we discussed that although circuit cutting is often shown to improve the quality of outcome, the exponential scaling of the required circuit executions with the number of cuts makes it largely impractical for bigger circuits. In this section, we show the use of operator backpropagation to scale circuit cutting. Since OBP can reduce the depth of the circuit, naturally the number of cuts required to partition the backpropagated circuit is expected to be lower. Therefore, it is possible to (i) use circuit cutting with lower budget for a given circuit, or (ii) scale circuit cutting to bigger and deeper circuits with the same budget. In this and future contexts, \emph{budget} should be considered to be directly proportional to the number of circuit executions. Moreover, when cutting is associated with a circuit without backpropagation -- we shall refer to it as \emph{vanilla cutting}.

\begin{figure}[htb]
    \begin{subfigure}[b]{0.9\textwidth}
        \centering
        \includegraphics[width=1.13\textwidth]{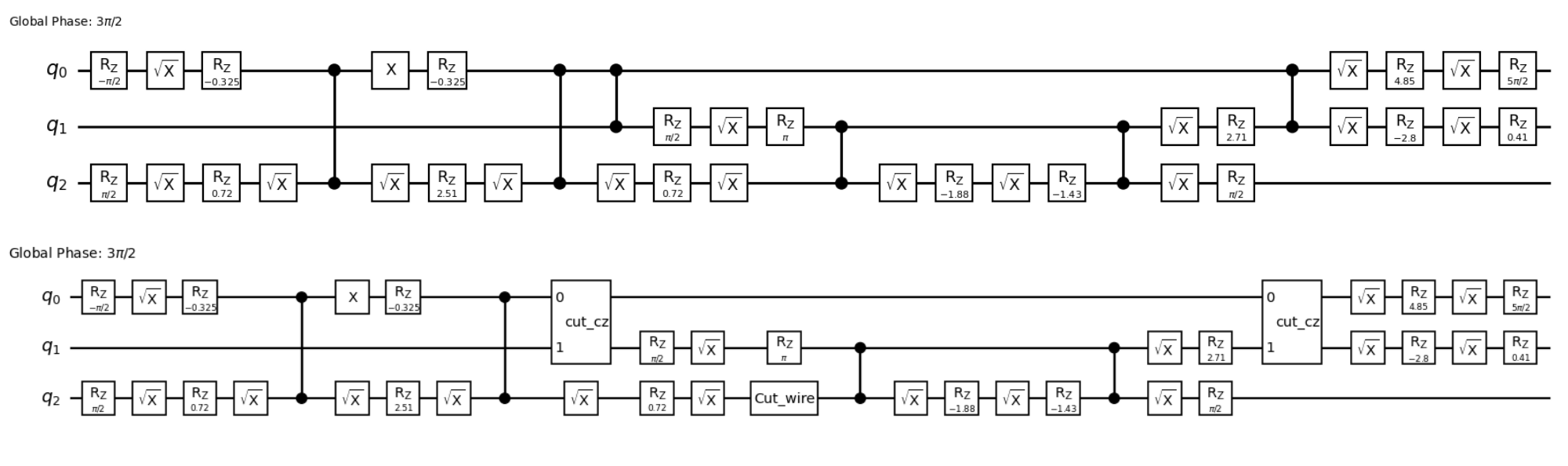}
        \caption{The original (top) circuit can be partitioned into 2 subcircuits (bottom) using 3 cuts (1 wire cut and 2 gate cuts)}
        \label{fig:subqaoa1}
    \end{subfigure}
    \hfill
    \begin{subfigure}[b]{0.9\textwidth}
        \centering
        \includegraphics[width=1.13\textwidth]{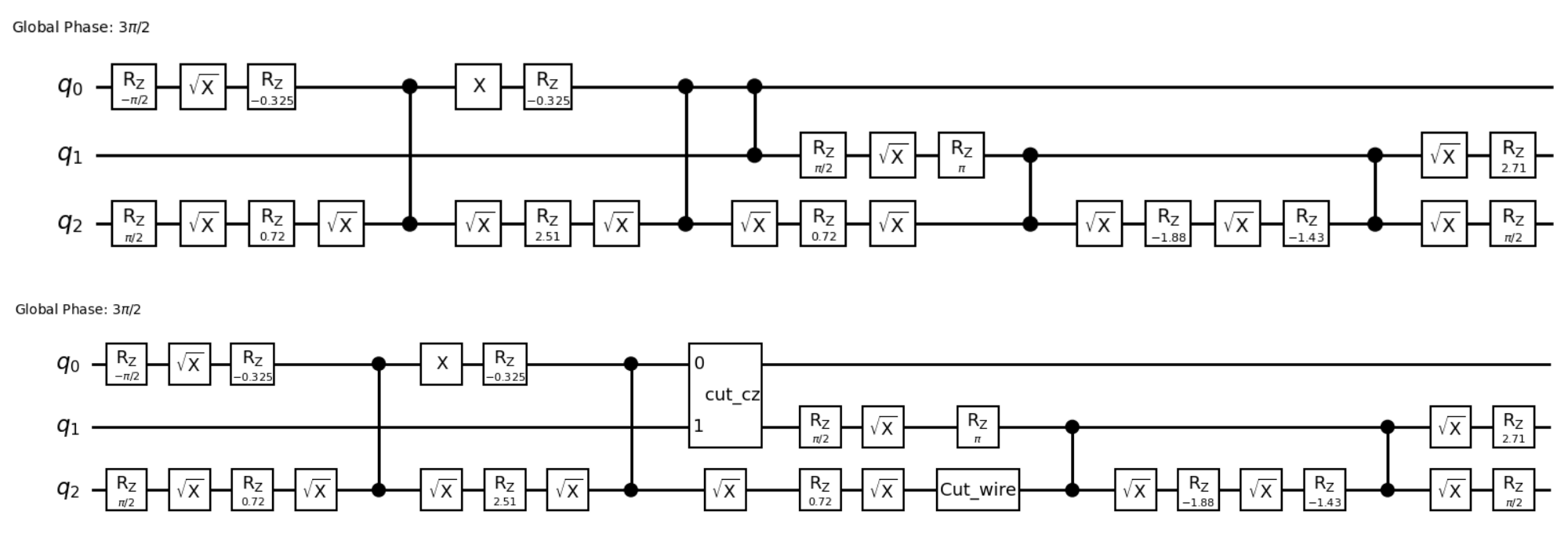}
        \caption{The backpropagated (top) circuit can be partitioned into 2 subcircuits (bottom) using 2 cuts (1 wire cut and 1 gate cut)}
        \label{fig:subqaoa2}
    \end{subfigure}

    \caption{The number of cuts required for a 3-qubit circuit for disjoint partitioning vs that required when OBP is used before cutting.}
    \label{fig:mainqaoa}
\end{figure}

On the other hand, OBP itself can increase the number of circuit executions by evolving the observable in such a way that it has more non-commuting terms. In Fig.~\ref{fig:mainqaoa} we show an example where the original circuit would require 3 cuts (2 gate cuts and 1 wire cut) for partitioning. Taking into account a simple observable $O = \frac{1}{n}\sum_{i=0}^{n-1} Z_i$, which can be evaluated with a single circuit execution, the overall circuit executions arising from circuit cutting are $9^2 \times 16 = 1296$. However, upon performing OBP, the circuit now requires only 2 cuts (1 gate cut and 1 wire cut), but the observable evolves to $0.3136761 IZI - 0.04732369 IIZ + 0.33333333 ZII - 0.11277595 IXZ - 0.32995694 IZX$ which has non-commuting terms and thus requires two circuit executions. Therefore, the total number of circuit executions required for cutting with OBP is $2 \times (9 \times 16) = 288$, leading to a $\sim 4\times$ reduction in overhead. Note that there is an exponential reduction in circuit executions from $9^2$ to $9$. However, there is a multiplicative increase in the number of circuits by $2$ due to the increase in non-commuting terms in the observable.

In general, let us consider a circuit $C$ and observable $O$ which can be partitioned into $s$ subcircuits using $k$ cuts. When a circuit is partitioned into $s$ subcircuits, the portion of the observable $O$ associated with each subcircuit is called its subobservable. For example, if $O = \otimes_{i=1}^s O_i$, then the subobservable $O_i$ is associated with subcircuit $i$. Now the number of non-commuting groups in $O_i$ will dictate the number of circuit executions required for calculating the expectation value. If $\eta_i$ be the number of circuit executions required for subcircuit $i$ (which will depend on the number and type of cuts associated with that subcircuit), and $g_i$ be the number of non-commuting groups in the subobservable associated with subcircuit $i$, then the total circuit executions required for vanilla cutting is $\sum_{i=1}^s g_i \eta_i$. 

We use the superscript $bp$ to denote the subcircuits and subobservables when the circuit has been backpropagated before cutting. The total number of circuit executions required for cutting equipped with OBP is $\sum_{i=1}^{s^{bp}} g_i^{bp}\eta_i^{bp}$. Using OBP with cutting is beneficial when

\begin{equation*}
    \label{eq: inequality}
    \sum_{i=1}^{s^{bp}} g_i^{bp}\eta_i^{bp} \leq \sum_{i=1}^s g_i \eta_i.
\end{equation*}



\subsection{Circuit cutting with OBP for VQE}
\label{subsec:vqe}

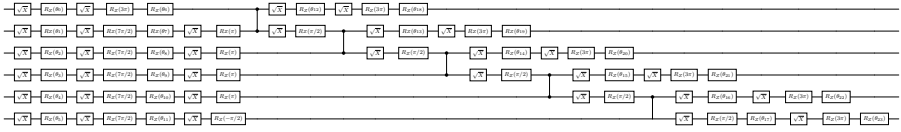
\begin{figure*}[htb]
    \centering

\resizebox{\textwidth}{!}{
\begin{quantikz}
    & \gate{\sqrt{X}} & \gate{R_Z(\theta_0)} & \gate{\sqrt{X}} & \gate{R_Z(3\pi)} & \gate{R_Z(\theta_6)} & & & \ctrl{1} & \gate{\sqrt{X}} & \gate{R_Z(\theta_{12})} & \gate{\sqrt{X}} & \gate{R_Z(3\pi)} & \gate{R_Z(\theta_{18})} & & & & & & & & & & & & & &\\
    & \gate{\sqrt{X}} & \gate{R_Z(\theta_1)} & \gate{\sqrt{X}} & \gate{R_Z(7\pi/2)} & \gate{R_Z(\theta_7)} & \gate{\sqrt{X}} & \gate{R_Z(\pi)} & \control{} & \gate{\sqrt{X}} & \gate{R_Z(\pi/2)} & \ctrl{1} & \gate{\sqrt{X}} & \gate{R_Z(\theta_{13})} & \gate{\sqrt{X}} & \gate{R_Z(3\pi)} & \gate{R_Z(\theta_{19})} & & & & & & & & & & & \\
    & \gate{\sqrt{X}} & \gate{R_Z(\theta_2)} & \gate{\sqrt{X}} & \gate{R_Z(7\pi/2)} & \gate{R_Z(\theta_8)} & \gate{\sqrt{X}} & \gate{R_Z(\pi)} & & &  & \control{} & \gate{\sqrt{X}} & \gate{R_Z(\pi/2)} & \ctrl{1} & \gate{\sqrt{X}} & \gate{R_Z(\theta_{14})} & \gate{\sqrt{X}} & \gate{R_Z(3\pi)} & \gate{R_Z(\theta_{20})} & & & & & & & & \\
    & \gate{\sqrt{X}} & \gate{R_Z(\theta_3)} & \gate{\sqrt{X}} & \gate{R_Z(7\pi/2)} & \gate{R_Z(\theta_9)} & \gate{\sqrt{X}} & \gate{R_Z(\pi)} & & & & & & & \control{} & \gate{\sqrt{X}} & \gate{R_Z(\pi/2)} & \ctrl{1} & \gate{\sqrt{X}} & \gate{R_Z(\theta_{15})} & \gate{\sqrt{X}} & \gate{R_Z(3\pi)} & \gate{R_Z(\theta_{21})} & & & & &\\
    & \gate{\sqrt{X}} & \gate{R_Z(\theta_4)} & \gate{\sqrt{X}} & \gate{R_Z(7\pi/2)} & \gate{R_Z(\theta_{10})} & \gate{\sqrt{X}} & \gate{R_Z(\pi)} & & & & & & & & & & \control{} & \gate{\sqrt{X}} & \gate{R_Z(\pi/2)} & \ctrl{1} & \gate{\sqrt{X}} & \gate{R_Z(\theta_{16})} & \gate{\sqrt{X}} & \gate{R_Z(3\pi)} & \gate{R_Z(\theta_{22})} & & \\
    & \gate{\sqrt{X}} & \gate{R_Z(\theta_5)} & \gate{\sqrt{X}} & \gate{R_Z(7\pi/2)} & \gate{R_Z(\theta_{11})} & \gate{\sqrt{X}} & \gate{R_Z(-\pi/2)} & & & & & & & & & & & & & & \gate{\sqrt{X}} & \gate{R_Z(\pi/2)} & \gate{R_Z(\theta_{17})} & \gate{\sqrt{X}} & \gate{R_Z(3\pi)} & \gate{R_Z(\theta_{23})} &  
\end{quantikz}
}
\caption{A 6-qubit EfficientSU2 circuit}
    \label{fig:vqe}
\end{figure*}

For VQE, we have considered a 6-qubit Efficient SU2 circuit (Fig.~\ref{fig:vqe}) with $O = \frac{1}{6} \sum_{i=0}^5 Z_i$ as the observable. We started the variational process with a random initial set of parameters, and executed upto convergence using the \textsc{cobyla} optimizer. The same initial set of parameters was used for the original (uncut) circuit, circuit equipped with cutting, and improving cutting with OBP. For OBP, the max\_qwc\_groups was set to $4$. The rationale behind this choice will be discussed in Sec.~\ref{sec:sa}. The number of cuts for vanilla cutting and OBP cutting, as provided by the \emph{automatic\_cut\_finder} of Qiskit-addon-cutting \cite{qiskit-addon-cutting} are 3 and 2 respectively.

\begin{figure}[htb]
\begin{center}
\includegraphics[scale=0.4]{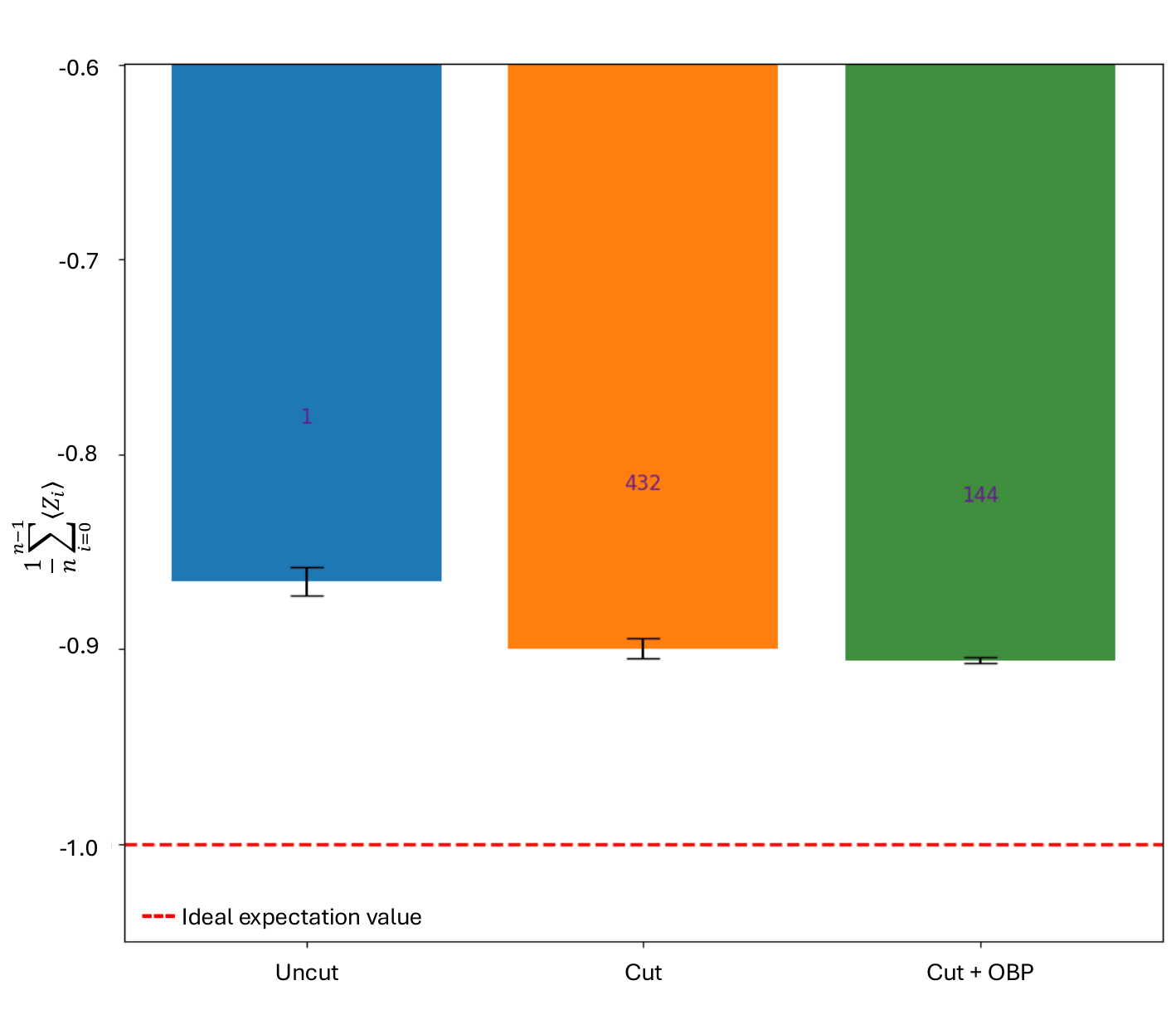}
\caption{A comparative analysis of quantum resource overhead in a 6-qubit VQE for the original (uncut) circuit, vs vanilla circuit cutting (cut) and improving circuit cutting with OBP (Cut+OBP). The values inside the bars indicate the number of quantum circuit executions required for each case.}
\label{fig:vqe_subexps_graph}
\end{center}
\end{figure}

Fig.~\ref{fig:vqe_subexps_graph} shows the expectation values of the uncut, vanilla circuit cutting, and cutting with OBP. We note that circuit cutting improves the quality of outcome over the uncut circuit as expected (and also observed in previous literature \cite{basu2022qer, khare2023parallelizing}). However, when equipped with OBP, circuit cutting shows a similar quality of outcome (in fact, slightly better) at $3\times$ lower resource. The number of circuit executions required for vanilla cutting and cutting with OBP are 432 and 144 respectively. Note that none of the experiments performed for this paper are equipped with any error mitigation or suppression techniques. Therefore, the performance improvement is due to circuit cutting and/or OBP only. Our results with VQE establishes the utility of OBP in attaining the quality of outcome provided by circuit cutting at a significantly lower resource requirement.

\subsection{Circuit cutting with OBP for Hamiltonian simulation}
\label{subsec:hs}

\begin{figure*}[htb]
\centering
\includegraphics[width=1\textwidth ]{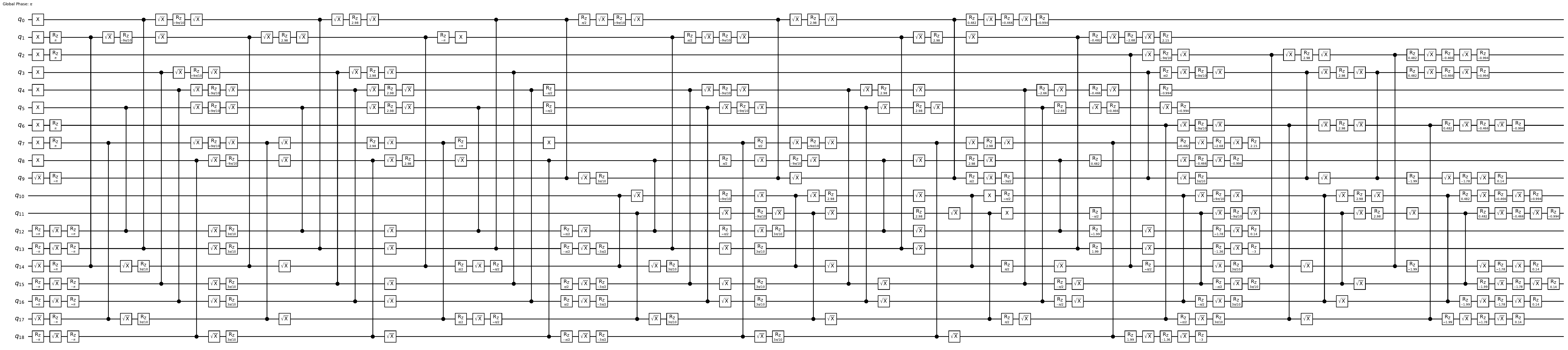}
\caption{A 19-qubit Hamiltonian simulation circuit}
    \label{fig:hs}

\end{figure*}

For Hamiltonian simulation circuit, we have considered a 19-qubit time evolution circuit (Fig.~\ref{fig:hs}) of a spin system for a time interval $t=0.2s$ governed by the Heisenberg XYZ Hamiltonian:
\begin{align*}
\hat{H} &= \sum_{(j,k)} \left( J_x X_j X_k + J_y Y_j Y_k + J_z Z_j Z_k \right) \\
&\quad + \sum_j \left( h_x X_j + h_y Y_j + h_z Z_j \right)
\end{align*}
where the external magnetic fields and the coupling constants are:
\begin{align*}
J_x &= \frac{\pi}{8}, \quad J_y = \frac{\pi}{4}, \quad J_z = \frac{\pi}{2}, \\
h_x &= \frac{\pi}{3}, \quad h_y = \frac{\pi}{6}, \quad h_z = \frac{\pi}{9}.
\end{align*}

Qubit interactions are defined by a bidirectional heavy-hex coupling map. $O = \frac{1}{6} \sum_{i=0}^5 Z_i$ is taken as the observable with trotterization as ``Lie Trotter" and trotter step 1. The same observable was used for the original (uncut) circuit, circuit equipped with cutting, and improving cutting with OBP. For OBP, the max\_qwc\_groups was set to $18$. The number of cuts for vanilla cutting and OBP cutting, as provided by the \emph{automatic\_cut\_finder} of Qiskit-addon cutting \cite{qiskit-addon-cutting} are 3 and 1 respectively.





\begin{figure}[htb]
\begin{center}
\includegraphics[scale=0.4]{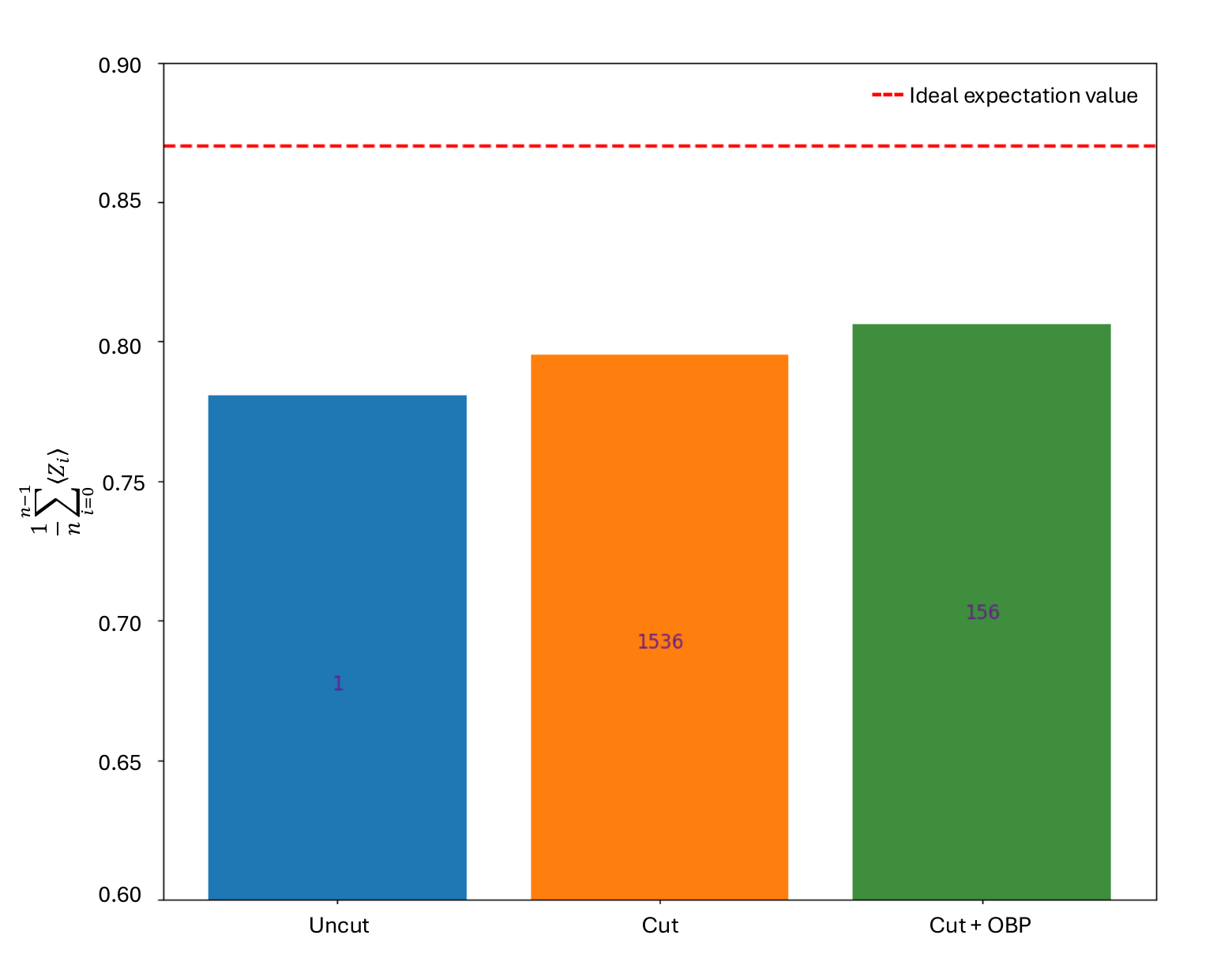}
\caption{Comparative analysis of quantum resource overhead in a 19-qubit Hamiltonian circuit for the original (uncut) circuit, vs using circuit cutting (cut) and improving cutting with OBP (Cut+OBP). The values inside the bars indicate the number of quantum circuit executions.}
\label{fig:hamiltonian_subexps_graph}
\end{center}
\end{figure}

Fig.~\ref{fig:hamiltonian_subexps_graph} shows the expectation values of the uncut, vanilla circuit cutting, and cutting with OBP for the Hamiltonian simulation circuit. We note from Fig.~\ref{fig:hamiltonian_subexps_graph} that circuit cutting improves the quality of outcome over the uncut circuit. However, when equipped with OBP, circuit cutting shows somewhat improved quality of outcome at $\sim 10\times$ lower resource. The number of circuit executions required for vanilla cutting and cutting with OBP are 1536 and 156 respectively. Our results with the Hamiltonian simulation circuit establishes the significant reduction in resource that can be obtained when cutting is equipped with OBP.


\subsection{Circuit cutting with OBP for multiple circuits from Benchpress}
\label{subsec:benchpress}

In this subsection, we are showing the reduction in the number of circuit executions obtained for different circuits from Benchpress \cite{nation2025benchmarking}. In the above two subsections we have already verified that we are getting better quality of outcomes when we are performing OBP on the quantum circuits before cutting. Therefore, we have not executed the Benchpress circuits on the hardware. Here, we are not going to discuss about the quality of the outcome; instead, we shall focus more on the reduction of circuit executions (i.e. quantum overhead). We have taken the same observable $O = \frac{1}{n}\sum_{i=0}^{n-1} Z_i$ for each Benchpress circuit to perform operator backpropagation (OBP) and cutting. Table ~\ref{tab:obp_summary} shows the number of circuit executions for vanilla cutting and cutting with OBP. In addition, we have varied the $max\_qwc\_groups$ from 2 to 5 to perform OBP on each circuit. 

Circuits having mid-circuit measurements have been removed from the dataset since these types of circuits do not, in general, conform to expectation value calculations, and it is not possible to backpropagate across a mid-circuit measurement. Moreover, circuits for which the number of cuts turned out to be 0 after backpropagation have also been removed from the dataset. Finally, for some circuits, OBP with $max\_qwc\_groups \in \{2, 3, 4, 5\}$ was able to backpropagate the entire circuit, thus eliminating the need for execution itself. These circuits are mainly clifford and were removed from the dataset as well.

\begin{table}[htb]
\centering
\caption{Implementing circuit cutting with OBP for multiple circuits from Benchpress \cite{nation2025benchmarking}}
\scriptsize
\resizebox{1\columnwidth}{!}{%
\begin{tabular}{|c|c|c|c|c|c|c|c|}
\hline
\multirow{2}{*}{Circuit} & \multirow{2}{*}{Qubits} & \multirow{2}{*}{Depth} 
& \multicolumn{2}{c|}{Vanilla cutting} 
& \multicolumn{3}{c|}{Cutting with OBP} \\
\cline{4-8}
& & & \ \# Cuts & \ \# Circuits & QWC & \ \# Cuts & \ \# Circuits \\
\hline
fredkin & 3 & 26 & 4 & 2592 & 2 & 3 & 1152 \\
\hline
qaoa & 3 & 27 & 3 & 576 & 2 & 2 & 192 \\
\hline
basis\_change & 3 & 53 & 4 & 2592 & 2 & 1 & 24 \\
\hline
linearsolver & 3 & 17 & 2 & 72 & 2 & 1 & 24 \\
\hline
small\_adder & 4 & 26 & 3 & 432 & 2 & 2 & 144 \\
\hline
vqe & 4 & 33 & 3 & 432 & 2 & 2 & 144 \\
\hline
qft & 4 & 37 & 4 & 2304 & 5 & 2 & 144 \\
\hline
state-encoded circuit ansatz & 11 & 156 & 4 & 12288 & 5 & 2 & 576 \\
\hline
adder & 18 & 319 & 2 & 72 & 2 & 1 & 18 \\
\hline
\rowcolor{red!20!white}
knn & 67 & 572 & 2 & 192 & 3 & 2 & 320 \\
\hline
big\_adder & 118 & 1200 & 2 & 108 & 2 & 1 & 84 \\
\hline
\end{tabular}
}
\label{tab:obp_summary}
\end{table}

We understand that Table \ref{tab:obp_summary} contains results of some circuits that do not conform to expectation value calculation e.g. adder; however, these circuits' results were included since it will prove the universality of our method. Here, we observe that the number of circuit executions for the KNN circuit increases after performing OBP with $max\_qwc\_groups \in \{2,3,4,5\}$ for cutting than vanilla cutting unlike the others. In section \ref{sec:sa}, we shall address this issue.

\section{Selection of optimal max\_qwc\_group}
\label{sec:select_qwc}

From our analysis of the BenchPress circuits, we observed that determining the optimal number of non-commuting groups ($max\_qwc\_groups$) to allow during operator backpropagation (OBP) is nontrivial. Using brute force method, with a few small values of $max\_qwc\_groups$, is not the ideal approach. We showed this for 67-qubit KNN, 19-qubit Hamiltonian simulation, and also observed this for many other circuits. This highlights the importance of carefully and accurately selecting the number of non-commuting groups permitted in OBP. Without this consideration, OBP may inadvertently introduce extra overhead instead of reducing the computational cost.

We further analyzed the relationship between the value of $max\_qwc\_groups$, and the total number of quantum circuit executions for cutting with OBP by visual representation in Fig.~\ref{fig:main} for the VQE and Hamiltonian simulation circuits. The figure clearly indicates the lack of any consistent or predictable pattern. This strongly highlights the need for a method that can determine the optimal value of $max\_qwc\_groups$ to be used in OBP, in order to achieve optimal results with minimal execution budget.

For this, we use a method called simulated annealing, the details of which are discussed in the following subsection.

\subsection{Simulated annealing to find max\_qwc\_groups}
\label{sec:sa}

\begin{figure}[htb]
    \centering

    \begin{subfigure}[b]{0.45\textwidth}
        \centering
        \includegraphics[width=\textwidth]{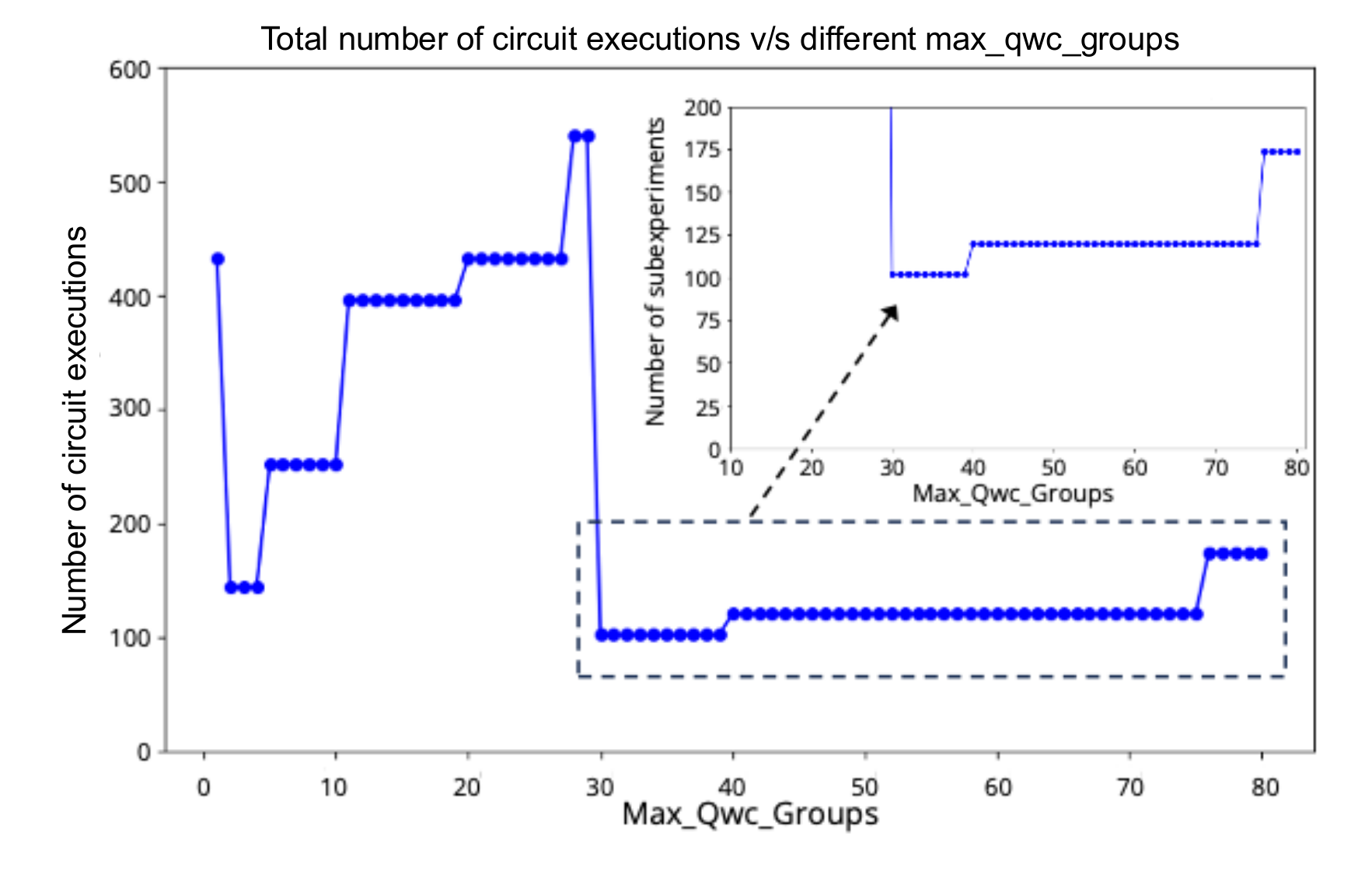}
        \caption{Number of circuit executions with different $max\_qwc\_groups$ for a 6-qubit Efficient SU2 circuit}
        \label{fig:sub1_non_determining_qwc_graph}
    \end{subfigure}
    \hfill
    \begin{subfigure}[b]{0.45\textwidth}
        \centering
        \includegraphics[width=\textwidth]{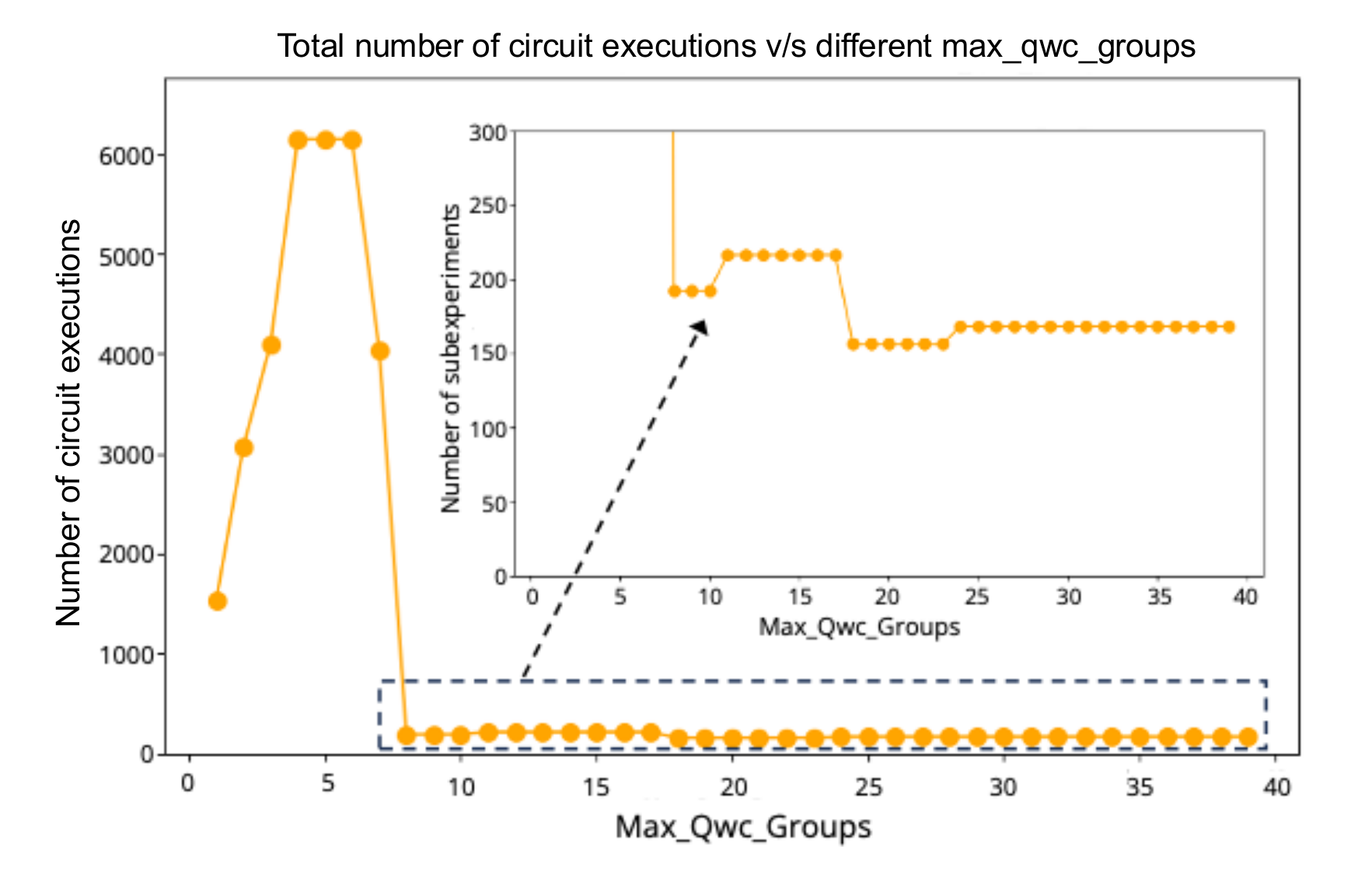}
        \caption{Number of circuit executions with different $max\_qwc\_groups$ for a 19-qubit Hamiltonian simulation circuit}
        \label{fig:sub2}
    \end{subfigure}

    \caption{Number of circuit executions required for cutting with OBP for different values of $max\_qwc\_groups$. These two plots show that there does not exist a simple relation between the value of $max\_qwc\_groups$ and the number of circuit executions.}
    \label{fig:main}
\end{figure}

In the previous section, we used $max\_qwc\_groups$ to be $4$ and $18$ for the VQE and Hamiltonian simulation problems respectively, and postponed the rationale for the same. We also noted that naive selection of the $max\_qwc\_groups$ does not always lead to improvement (e.g. 67-qubit KNN circuit). In this section, we discuss the selection of the $max\_qwc\_groups$ for a given circuit and observable. A general method for this can be to decide a range $\{1, 2, \hdots, w_k\}$ of the $max\_qwc\_groups$, try backpropagation for each of them, and take the one with the minimum circuit executions. However, it should be noted that backpropagation comes at the cost of classical overhead. Therefore, sequentially trying out many different values will also lead to significant classical overhead. Moreover, in Fig.~\ref{fig:main} (a) and (b) we show the number of circuit executions for different $max\_qwc\_groups$ for the VQE and Hamiltonian simulation examples used in Sec.~\ref{sec:obp_cut} respectively. It is obvious from this figure that there is no immediate pattern which can be exploited to decide the optimal $max\_qwc\_group$ for a given circuit and observable.

\begin{algorithm}[htb]
\caption{Objective function}
\label{alg:obj_fun}
    \begin{algorithmic}[1]
    \REQUIRE circuit, $w \in \mathbb{I}$
\ENSURE $num\_circuits$
        \STATE backpropagated\_circuit $\leftarrow$ backpropagate (circuit, $w$)
            \IF {the entire circuit is backpropagated}
                \STATE $num\_circuits = 0$
            \ELSE
                \STATE cut the backpropagated\_circuit
                \STATE $num\_circuits=$ executions required for cutting the backpropagated circuit
            \ENDIF
    \end{algorithmic}
\end{algorithm}

\begin{algorithm}[htb]
\caption{Finding the optimal $max\_qwc\_groups$ $w_{opt}$}
\label{alg:sa}
\begin{algorithmic}[1]
\REQUIRE circuit, $bound_{upper}$, $bound_{lower}$, $num_{iter}$, $step\_size$, initial temperature $T$.
\ENSURE $w_{opt}, opt\_num\_circuits$
\STATE $w_{opt} = \text{random integer} \in \{bound_{lower}, bound_{upper}\}$
\STATE $opt\_num\_circuits =$ Objective Function(circuit, $w_{opt})$ calculated using Algorithm~\ref{alg:obj_fun}
\STATE $eval\_dict[w_{opt}] = opt\_num\_circuits$
\STATE $iters \leftarrow 0$
\STATE $w, num\_circuits \leftarrow w_{opt}, opt\_num\_circuits$
\WHILE{$iters \leq num\_iters$}
    \STATE $w_{new} = \text{random integer} \in \{w_{opt} - step\_size, w_{opt} + step\_size\}$
    \IF{$w_{new} < bound_{lower}$}
        \STATE $w_{new} = bound_{lower}$
    \ENDIF
    \IF{$w_{new} > bound_{upper}$}
        \STATE $w_{new} = bound_{upper}$
    \ENDIF
    \IF{$w_{new} \in eval\_dict$}
        \STATE $num\_circuits\_new = eval\_dict[w_{new}]$
    \ELSE
        \STATE $num\_circuits\_new =$ Objective Function(circuit, $w_{new})$ calculated using Algorithm~\ref{alg:obj_fun}
        \STATE $eval\_dict[w_{new}] = num\_circuits\_new$
    \ENDIF
    \IF{$num\_circuits\_new < opt\_num\_circuits$}
        \STATE $opt\_num\_circuits, w_{opt} \leftarrow num\_circuits\_new, w_{new}$
    \ENDIF
    \IF{$num\_circuits\_new < num\_circuits$}
        \STATE $w \leftarrow w_{new}$
    \ELSE
        \STATE $w \leftarrow w_{new}$ with probability $p = exp(-\frac{num\_circuits\_new - num\_circuits}{T})$
    \ENDIF
    \STATE $T \leftarrow \frac{T}{num\_iter + 1}$
    \STATE $num\_iter = num\_iter + 1$
\ENDWHILE
\end{algorithmic}
\end{algorithm}

Therefore, we resort to a heuristic technique called Simulated Annealing \cite{kirkpatrick1983optimization} to find a \emph{good} $max\_qwc\_group$ for a given circuit. Algorithm~\ref{alg:sa} used for this is the traditional Simulated Annealing algorithm. Here, the initial value used for $b_0$ is $1$, and we have used $b_1 = 40$. The value of $b_1$ can be modified as per the classical resource available. Moreover, we have maintained a dictionary $eval\_dict$ where we stored the values of $w$ (non-commuting groups) and its corresponding number of circuit executions for all $w$ traversed by the algorithm. This ensures that if the same value of $w$ is repeated during the process, the number of circuit executions is not computed multiple times, thus saving classical resources. For our experiment, we have set the number of maximum iterations of the algorithm to be $20$, and repeated the algorithm $5$ times in parallel using \emph{Qiskit Serverless} \cite{gambetta2024quantum}. This method can be further improved by maintaining a global dictionary $eval\_dict$, instead of a local copy for each thread, with message passing to each thread. This will further reduce the classical overhead to determine the number of circuit executions.

Using Algorithm~\ref{alg:sa}, which calls Algorithm~\ref{alg:obj_fun} to perform backpropagation using \textsc{qiskit-addon-obp} and determine the number of circuit executions using \textsc{qiskit-addon-cutting}, we obtained the $max\_qwc\_group$ to be $4$ for the 6-qubit VQE example, and $18$ for the Hamiltonian simulation example. Moreover, in Table~\ref{tab:obp_summary}, we noted that the 67-qubit KNN circuit actually ended up with more circuit executions when performing with OBP for $max\_qwc\_group \in \{2, 5\}$. Using the simulated annealing method, we obtained a number of circuit executions ($176$) for $max\_qwc\_group = 10$, which is lesser than the number of circuit executions ($192$) for vanilla cutting.

\subsection{Results for different observables}
\label{subsec:obs_diff}

In all the above experiments, although the circuit was changing, we fixed the observable to be $O = \frac{1}{n} \sum_{i=0}^{n-1} Z_i$. However, the improvement obtained will also vary with the observable. Here, we perform an experiment with the 67-qubit KNN for different observables of the form $Z^b$ for $b = \{1,2,3\}$. Here $b$ denotes the weight of the observable, and is averaged over all nearest-neighbor terms. For example, $b=1$ corresponds to the observable $O$ mentioned above. $b = 2$ implies the observable $O = \frac{1}{n-2} \sum_{i=0}^{n-2} Z_i Z_{i+1}$. In Table~\ref{tab:cutting_observables_summary} we show that the improvement is retained for different weight observables for the 67-qubit KNN circuit, with higher weight observables often providing even lower number of circuit executions. The similar trend was observed for other circuits and observables (beyond weight-1) as well, i.e., circuit cutting equipped with OBP, where the $max\_qwc\_group$ is carefully selected using simulated annealing, resulted in fewer circuit executions.


\begin{table}[H]
\centering
\caption{Number of circuit executions required for different observable weights in vanilla cutting v/s cutting with OBP using simulated annealing for the 67-qubit KNN circuit}
\scriptsize
\resizebox{1\columnwidth}{!}{%
\begin{tabular}{|c|c|c|c|}
\hline
\multirow{2}{*}{Observable Weight} 
& \multirow{2}{*}{Vanilla Cutting} 
& \multicolumn{2}{c|}{Cutting with OBP using SA} \\
\cline{3-4}
& & Optimal QWC Groups & \# Circuit Executions \\
\hline
Weight-1 & 192 & 10 & 176 \\
\hline
Weight-2 & 192 & 6 & 72 \\
\hline
Weight-3 & 192 & 6 & 72 \\
\hline
\end{tabular}
}
\label{tab:cutting_observables_summary}
\end{table}

The results in Table~\ref{tab:obp_summary} and ~\ref{tab:cutting_observables_summary} show that our proposed method of using OBP to lower the circuit cutting overhead holds across different circuits and observables as long as a good value of $max\_qwc\_group$ is selected. Therefore, this, together with our proposed simulated annealing based method to select the $max\_qwc\_group$, is applicable to any (circuit, observable) combination.

\section{Effect of truncation error}
\label{sec:truncation}
\begin{figure}[htb]
\begin{center}
\includegraphics[scale=0.35]{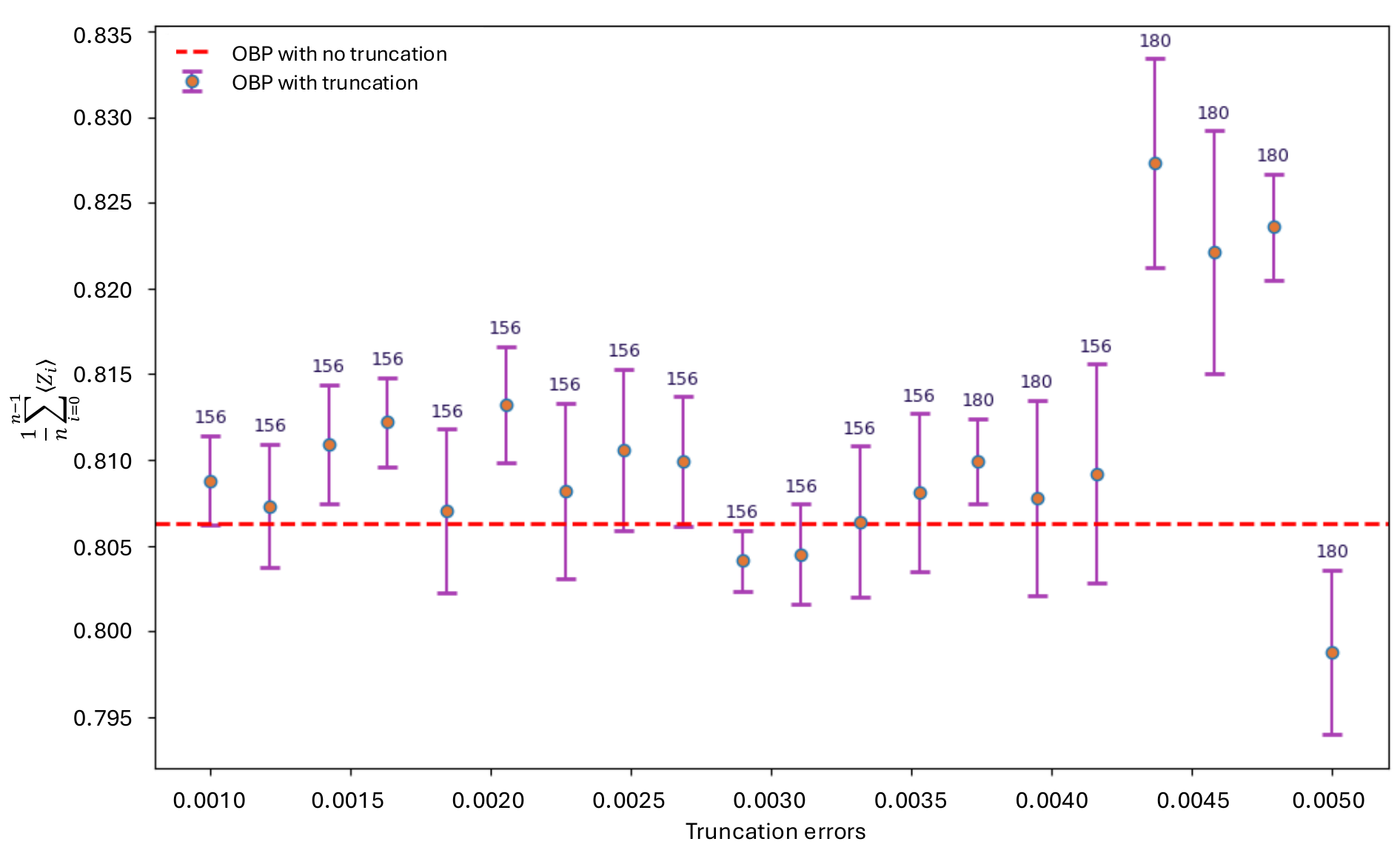}
\caption{The expectation value of a 19-qubit Hamiltonian simulation circuit for circuit cutting equipped with OBP with truncation}
\label{fig:trunc_graph}
\end{center}
\end{figure}

In all the experiments in the previous sections, we have considered backpropagation without truncation. When the number of non-commuting terms in the observable grows significantly with backpropagation, it is often a practice to remove some of the terms with low coefficients \cite{fuller2025improved}. This allows for more backpropagation respecting the circuit execution overhead, leading to an even shallower circuit, at the expense of some approximation error coming from the truncated terms. In this section, we study the effect of allowing different truncation errors in OBP for the 19-qubit Hamiltonian simulation circuit with the same $max\_qwc\_group = 18$ used in Sec.~\ref{subsec:hs}. The truncation is performed with a pre-defined upper bound on the error for each slice backpropagated. We selected 20 equally distributed values of truncation errors from $0.001$ to $0.005$, and calculated the expectation value for the observable $O = \frac{1}{n} \sum_{i=0}^{n-1} Z_i$ for cutting equipped with OBP with truncation. In Fig.~\ref{fig:trunc_graph} we show the mean and standard deviation of the expectation value obtained when the experiment was repeated 10 times for each truncation error. 

We note that there does not seem to be a trivial relation between the amount of truncation with the quality of outcome. For example, we obtained the best value within our selected range for a value of truncation error, which required $180$ circuit executions. For some truncation errors, the expectation value for the observable $O = \frac{1}{n} \sum_{i=0}^{n-1} Z_i$ after backpropagation is better than that without truncation, although the number of circuit executions increases from $156$ to $180$. On the other hand, for values of truncation errors where the number of circuit execution is still $156$, the result seems to yield a poorer result than backpropagation without truncation most of the time.

This indication, that approximation error in each subcircuit amplifies the overall error after reconstruction, was observed in \cite{majumdar2022error}. However, that alone does not explain why lower truncation error leads to poorer performance at times. This most likely has to do with the facts that - (i) truncation does not always necessarily lead to lower depth circuit, and (ii) the approximation error induced by truncation need not always be overshadowed by the reduction in noise due to deeper backpropagation. Minimizing the number of circuit executions was a good metric for backpropagation without truncation, where the backpropagated circuit was functionally equivalent to the original one. However, in the presence of truncation, where functional equivalence is not preserved, the number of circuit executions alone cannot be the ideal metric. We postpone the study of circuit cutting along with backpropagation with truncation for future endeavor.


\section{Discussion: Using Multi-product formulas with circuit cutting}
\label{sec:discussion}
In this paper we discussed using OBP to reduce the overhead of circuit cutting. The primary utility of OBP was to reduce the depth of the circuit, thereby reducing the number of cuts required for disjoint partitioning of the circuit. Another method, which can reduce the depth of the circuit is Multi-product formulas (MPF) \cite{zhuk2024trotter, vazquez2023well}. This method provides high quality simulation of Hamiltonian time evolution using fewer trotter steps at the expense of more circuit executions. The trotter error of a Hamiltonian time evolution lowers with the number of trotter steps $r$, reaching $0$ as $r \rightarrow \infty$. Increasing the number of trotter steps naturally increases the depth of the circuit. Using MPF, it is possible to obtain the trotter error of $r$ steps using $r' < r$ steps, by taking a weighted average of multiple circuit executions. Naturally, the circuit with fewer trotter steps $r'$ is expected to require fewer cuts for disjoint partitioning. 

Therefore, it is also possible to use MPF independently or together with OBP to reduce the depth, and hence the number of cuts required for disjoint partitioning, of a Hamiltonian simulation circuit. However, since both MPF and OBP increase the number of circuit executions, the overhead arising from this should also be considered to avoid eventually increasing the number of circuit executions instead of reducing it (recall Fig.~\ref{fig:mainqaoa} and the corresponding discussion). 

Given the similarity in the circuital nature -- both OBP and MPF lead to increased number of circuit execution, where each circuit is of lower depth, a similar analysis, as shown in this paper, is likely to work for MPF as well. However, unlike OBP, which is applicable for any circuit, MPF is applicable only for time evolution of Hamiltonian, making its usability highly limited. Therefore, we abstain from a more detailed study of the scalability of circuit cutting using MPF with or without OBP.

\section{Conclusion}
\label{sec:conclusion}

Circuit cutting, although a widely studied tool to reduce the noise in the system, suffers from the scalability issue since the number of circuit executions increase exponentially with the number of cuts. This limits the applicability of circuit cutting for practical circuits. In this paper, we show that it is possible to significantly lower the overhead of circuit cutting when it is equipped with operator backpropagation. Our results show a $3\times$ and $10\times$ reduction in the number of circuit executions for VQE and Hamiltonian simulation respectively. We tested our method for multiple other circuits from the \emph{Benchpress} dataset, and for observables of different weights, and found this method to work for all of them unanimously. The major issue in using OBP effectively with circuit cutting is to select the optimal $max\_qwc\_group$ for OBP. We show that naive selection of this parameter can eventually end up increasing the number of circuit executions. To overcome this, we proposed the use of simulated annealing to determine a good value for this parameter, and obtained lower number of circuit executions for all scenarios when compared with vanilla circuit cutting. Our proposed method opens up a path to make circuit cutting feasible for deeper and wider circuits, and increases its applicability beyond the current scenario.

\section*{Acknowledgement}
The authors acknowledge useful discussions with Nathan Earnest-Noble, Bryce Fuller, Caleb Johnson and Ibrahim Shehzad.

\bibliographystyle{spmpsci} 
\bibliography{sample}

\begin{thebibliography}{10}
\providecommand{\url}[1]{{#1}}
\providecommand{\urlprefix}{URL }
\expandafter\ifx\csname urlstyle\endcsname\relax
  \providecommand{\doi}[1]{DOI~\discretionary{}{}{}#1}\else
  \providecommand{\doi}{DOI~\discretionary{}{}{}\begingroup \urlstyle{rm}\Url}\fi

\bibitem{abbas2024challenges}
Abbas, A., Ambainis, A., Augustino, B., B{\"a}rtschi, A., Buhrman, H., Coffrin, C., Cortiana, G., Dunjko, V., Egger, D.J., Elmegreen, B.G., et~al.: Challenges and opportunities in quantum optimization.
\newblock Nature Reviews Physics pp. 1--18 (2024)

\bibitem{alexeev2024quantum}
Alexeev, Y., Amsler, M., Barroca, M.A., Bassini, S., Battelle, T., Camps, D., Casanova, D., Choi, Y.J., Chong, F.T., Chung, C., et~al.: Quantum-centric supercomputing for materials science: A perspective on challenges and future directions.
\newblock Future Generation Computer Systems \textbf{160}, 666--710 (2024)

\bibitem{basu2023towards}
Basu, S., Born, J., Bose, A., Capponi, S., Chalkia, D., Chan, T.A., Doga, H., Flother, F.F., Getz, G., Goldsmith, M., et~al.: Towards quantum-enabled cell-centric therapeutics.
\newblock arXiv preprint arXiv:2307.05734  (2023)

\bibitem{basu2022qer}
Basu, S., Saha, A., Chakrabarti, A., Sur-Kolay, S.: i-qer: An intelligent approach towards quantum error reduction.
\newblock ACM Transactions on Quantum Computing \textbf{3}(4), 1--18 (2022)

\bibitem{beguvsic2025simulating}
Begu{\v{s}}i{\'c}, T., Hejazi, K., Chan, G.K.: Simulating quantum circuit expectation values by clifford perturbation theory.
\newblock The Journal of Chemical Physics \textbf{162}(15) (2025)

\bibitem{bhoumik2023distributed}
Bhoumik, D., Majumdar, R., Saha, A., Sur-Kolay, S.: Distributed scheduling of quantum circuits with noise and time optimization.
\newblock arXiv preprint arXiv:2309.06005  (2023)

\bibitem{qiskit-addon-cutting}
Bra\'{n}czyk, A.M., {Carrera Vazquez}, A., Egger, D.J., Fuller, B., Gacon, J., Garrison, J.R., Glick, J.R., Johnson, C., Joshi, S., Pednault, E., Pemmaraju, C.D., Rivero, P., Shehzad, I., Woerner, S.: {Qiskit addon: circuit cutting}.
\newblock \url{https://github.com/Qiskit/qiskit-addon-cutting} (2024).
\newblock \doi{10.5281/zenodo.7987997}

\bibitem{brenner2023optimal}
Brenner, L., Piveteau, C., Sutter, D.: Optimal wire cutting with classical communication.
\newblock arXiv preprint arXiv:2302.03366  (2023)

\bibitem{di2024quantum}
Di~Meglio, A., Jansen, K., Tavernelli, I., Alexandrou, C., Arunachalam, S., Bauer, C.W., Borras, K., Carrazza, S., Crippa, A., Croft, V., et~al.: Quantum computing for high-energy physics: State of the art and challenges.
\newblock Prx quantum \textbf{5}(3), 037001 (2024)

\bibitem{fuller2025improved}
Fuller, B., Tran, M.C., Lykov, D., Johnson, C., Rossmannek, M., Wei, K.X., He, A., Kim, Y., Vu, D., Sharma, K., et~al.: Improved quantum computation using operator backpropagation.
\newblock arXiv preprint arXiv:2502.01897  (2025)

\bibitem{gambetta2024quantum}
Gambetta, J.: Quantum software for the utility-scale and beyond.
\newblock In: IEEE International Conference on Quantum Computing and Engineering (2024)

\bibitem{khare2023parallelizing}
Khare, T., Majumdar, R., Sangle, R., Ray, A., Seshadri, P.V., Simmhan, Y.: Parallelizing quantum-classical workloads: Profiling the impact of splitting techniques.
\newblock In: 2023 IEEE International Conference on Quantum Computing and Engineering (QCE), vol.~1, pp. 990--1000. IEEE (2023)

\bibitem{kirkpatrick1983optimization}
Kirkpatrick, S., Gelatt~Jr, C.D., Vecchi, M.P.: Optimization by simulated annealing.
\newblock science \textbf{220}(4598), 671--680 (1983)

\bibitem{majumdar2022error}
Majumdar, R., Wood, C.J.: Error mitigated quantum circuit cutting.
\newblock arXiv preprint arXiv:2211.13431  (2022)

\bibitem{mitarai2021constructing}
Mitarai, K., Fujii, K.: Constructing a virtual two-qubit gate by sampling single-qubit operations.
\newblock New Journal of Physics \textbf{23}(2), 023021 (2021)

\bibitem{mitarai2021overhead}
Mitarai, K., Fujii, K.: Overhead for simulating a non-local channel with local channels by quasiprobability sampling.
\newblock Quantum \textbf{5}, 388 (2021)

\bibitem{nation2025benchmarking}
Nation, P.D., Saki, A.A., Brandhofer, S., Bello, L., Garion, S., Treinish, M., Javadi-Abhari, A.: Benchmarking the performance of quantum computing software for quantum circuit creation, manipulation and compilation.
\newblock Nature Computational Science pp. 1--9 (2025)

\bibitem{peng2020simulating}
Peng, T., Harrow, A.W., Ozols, M., Wu, X.: Simulating large quantum circuits on a small quantum computer.
\newblock Physical review letters \textbf{125}(15), 150504 (2020)

\bibitem{saleem2021divide}
Saleem, Z.H., Tomesh, T., Perlin, M.A., Gokhale, P., Suchara, M.: Divide and conquer for combinatorial optimization and distributed quantum computation.
\newblock arXiv preprint arXiv:2107.07532  (2021)

\bibitem{schmitt2025cutting}
Schmitt, L., Piveteau, C., Sutter, D.: Cutting circuits with multiple two-qubit unitaries.
\newblock Quantum \textbf{9}, 1634 (2025)

\bibitem{tang2021cutqc}
Tang, W., Tomesh, T., Suchara, M., Larson, J., Martonosi, M.: Cutqc: using small quantum computers for large quantum circuit evaluations.
\newblock In: Proceedings of the 26th ACM International conference on architectural support for programming languages and operating systems, pp. 473--486 (2021)

\bibitem{vazquez2023well}
Vazquez, A.C., Egger, D.J., Ochsner, D., Woerner, S.: Well-conditioned multi-product formulas for hardware-friendly hamiltonian simulation.
\newblock Quantum \textbf{7}, 1067 (2023)

\bibitem{zhuk2024trotter}
Zhuk, S., Robertson, N.F., Bravyi, S.: Trotter error bounds and dynamic multi-product formulas for hamiltonian simulation.
\newblock Physical Review Research \textbf{6}(3), 033309 (2024)

\end{thebibliography}
\end{document}